\documentstyle[aps,preprint]{revtex}
\begin{document}
\draft

\title{STATISTICAL BINARY DECAY OF $^{35}$Cl + $^{24}$Mg \\
AT $\approx$ 8 MEV/NUCLEON}

\vskip 2.2 cm

\author{ R.~Nouicer, C.~Beck, D.~Mahboub, T.~Matsuse, B.~Djerroud,
R.M.~Freeman and A.~Hachem}

\vskip 1.0 cm

\address {\it Centre de Recherches Nucl\'eaires, Institut National de Physique
Nucl\'eaire et de Physique des Particules, Centre de la Recherche
Scientifique/Universit\'e Louis Pasteur, B.P.28, F-67037 Strasbourg Cedex~2,
France }

\vskip 2.0 cm

\author { Sl.~Cavallaro, E.~De Filippo, G.~Lanzan\'o , A.~Pagano and
M.L.~Sperduto }

\vskip 1.0 cm

\address {\it Dipartimento di Fisica dell'Universit\'a di Catania, INFN and LNS
Catania, I-95129 Catania, Italy }

\vskip 2.0 cm

\author {R.~Dayras, E.~Berthoumieux, R.~Legrain and E.~Pollacco}

\vskip 1.0 cm

\address{\it DAPNIA/SPhN, C.E. Saclay, F-91191 Gif sur Yvette Cedex, France }

\date{\today}
\maketitle

\newpage

\begin{abstract}

{\it The properties of the two-body channels in the $\displaystyle{^{35}{\rm
Cl} + {^{24}{\rm Mg}}} $ reaction at a bombarding energy of $\displaystyle{275\
{\rm MeV} }$, have been investigated by using fragment-fragment coincident
techniques. The exclusive data show that the majority of events arises from a
binary-decay process. The rather large number of secondary light
charged-particles emitted from the two excited exit fragments are consistent
with the expectations of the Extended Hauser-Feshbach Method. Therefore no
evidence  for the occurence of ternary break-up events is observed. }

\end{abstract}


\pacs{{\bf PACS} numbers: 25.70.-z, 24.60}

\newpage

The systematic trends of the fusion-fission (FF) process in the $\displaystyle{
 \rm 40 \leq A_{CN} \leq 60}$ mass region is at present well established
\cite{R1,R2}. In general, calculations using the transition-state model
formalism \cite{R1} reproduce the basic properties of the binary fragments
providing evidence for a FF mechanism.

\bigskip

In this note $\displaystyle{^{35}{\rm Cl}  + {^{24}{\rm Mg}}}$ experimental
coincident results are presented and discussed within an alternative model
\cite{R3}. The data have been obtained with a $\displaystyle{275\ {\rm MeV} }$
\ $\displaystyle{^{35}{\rm Cl}}$ beam (well above the FF energy threshold
\cite{R2}) provided by the Saclay Booster Tandem facility. The reaction
products were detected in singles mode between $\displaystyle{-45^{o}}$ and
$\displaystyle{85^{o}}$ and, in a coincident mode, between
$\displaystyle{-37^{o}}$ and $\displaystyle{95^{o}}$, by seven ionization
chambers, at a pressure of $\displaystyle{52\ {\rm torrs}}$ of
${\displaystyle{\rm CF_{4}}}$ gas, followed by a $\displaystyle{ 500 \ {\rm \mu
 m }}$ thick Si(SB) detector. On an event-by-event basis, corrections were
applied for energy loss in the target and window foils and for the pulse-height
defect in the Si detectors.

\bigskip

The elemental Z distribution of integrated fully-damped and evaporation
residues (ER) cross sections (points) extracted from the singles yields are
plotted in Fig.1 with statistical model calculations using the Extended
Hauser-Feshbach Method (EHFM) and EHFM+CASCADE \cite{R3,R4}. The total FF and
ER cross sections are ${\displaystyle{\rm  \sigma_{\rm FF} = 137 \pm \ 5 \ {\rm
mb } }}$ and ${\displaystyle{\rm  \sigma_{\rm ER} =\ 722 \  \pm  \ 197 \ {\rm
mb }}}$ respectively in excellent agreement with the previous inclusive study
of Cavallaro et al. \cite{R5}. EHFM assumes the fission decay width to be
proportional to the available phase space at the scission point. It starts with
the compound nucleus (CN) formation hypothesis and then follows the system
which is allowed to decay by first-chance binary fission or fusion-evaporation.
Subsequent light-particle emissions from the fully-accelerated fission
fragments are explicitly taken into account. The full procedure including
secondary emission is denoted as EHFM+CASCADE \cite{R3,R4}. For this
calculation the simple case without deformation has been chosen. The
diffuseness parameter is ${\displaystyle{{{\rm \Delta } \ = \ 1\ {\rm
\hbar}}}}$ and critical angular  momentum ${\displaystyle{ {\rm L_{crit}=44 \
{\rm \hbar }} }}$. The neck length parameter  of the scission-point
parametrization is ${\displaystyle{ {\rm d=3.8\ {\rm fm}}}}$ according to the
systematics \cite{R3}.

\newpage

In Fig.1 EHFM+CASCADE clearly shows that the effects of secondary particle
emission play an important role in the deexcitation scheme. The small
displacement between the calculations and  experimental points for ER's in the
Fig.1 may be due to the deformation of the ${\displaystyle{^{59} {\rm Cu}}}$ CN
\cite{R6}.

\bigskip

The experimental "missing charge distributions" extracted from the coincident
yields are shown in Fig.2 as solid histograms and compared to the predictions
of the full EHFM+CASCADE calculations which are displayed for chosen angle
settings as dashed histograms. The average missing charge ${ \displaystyle{
{\rm \Delta Z} }}$ which is around 4 charge units, in agreement  with the
recently established systematics \cite{R4,R7}, is most likely lost through
particle emission from either the excited composite system or a secondary
sequential evaporation from both binary-reaction partners. The small shoulder
which is observed in Fig.2 for ${ \displaystyle{ \rm Z=12}}$ is due to a C
contamination of the Mg target. This has been verified by studying the reaction
${ \displaystyle{ ^{35}{\rm Cl} + {^{12}{\rm C }} }}$ at ${ \displaystyle{ \rm
E_{lab} = \ 278 \,{\rm MeV}}}$ \cite{R4}.

\bigskip

The possible occurrence of ternary processes that involve three massive
fragments in competition with the binary-decay mechanisms has been searched
for. No strong evidence is observed for the onset of ternary processes in the
${ \displaystyle{ ^{35}{\rm Cl} + {^{24}{\rm Mg }} }}$ reaction data for
incident energies lower than 10 MeV/nucleon as recently claimed in
ref.\cite{R8} for more massive systems. More quantitatively the average
$\displaystyle{\rm  < Z_{1} + Z_{2} >  }$ calculated from the EHFM+CASCADE code
can be compared with the experimental data in Fig.3. The agreement between the
calculations and the data is quite satisfactory. Naturally this comparison is
only relevant for mechanisms assuming a full-energy damping as in the case of
CN and/or orbiting deep-inelastic processes. As a consequence the discrepancies
observed in Fig.3 in the region $\displaystyle{ \rm Z_{1} = }$ 13 \ to \ 16 may
indicate that these products have a different but still unknown origin.\\

\newpage

In summary, the comparison of fragment-fragment coincidence data with the
predictions of the Extended Hauser-Feshbach Method indicates that the majority
of events arises from a binary-decay process with rather large numbers of
secondary light charged-particles emitted from the two excited exit fragments.
No evidence for the occurence of a three-body processs is clearly observed at
this energy. In addition, the good agreement between the experiment and the
statistical model may indicate that pre-scission emission of light particles
from the dinuclear system might be weak in this mass region in contrast to
heavier reactions.

\newpage

\centerline{\bf ACKNOWLEDGEMENTS}

\vskip 2.0 cm

The authors wish to thank the Booster Tandem Service at Saclay for
the kind hospitality and their technical support.

%
%

\begin{figure}
\caption { $\displaystyle{^{35}{\rm Cl }+ {^{24}{\rm Mg}}}$ elemental
distribution (points) at 275 MeV compared with EHFM (dashed histograms) and
EHFM+CASCADE (solid histograms) calculations. The open circles are the
inclusive data measured at $\displaystyle{280 \ {\rm MeV [5]}.}$}
\label{FIG.1 :}
\end{figure}

\vskip 1.0 cm

\begin{figure}
\caption{ The missing charge distributions (solid histograms) as measured for
$\displaystyle{{{\rm ^{35}Cl} + {^{24}{\rm Mg}}}}$ reaction for each charge
$\displaystyle{ {\rm (Z_{1} = 5 \ to \ 12 )}}$. The dashed histograms are
EHFM+CASCADE calculations under the same conditions ${ \displaystyle{\rm
Z_{1,2} \geq \, 5}}.$}
\label {FIG.2 : }
\end{figure}

\vskip 1.0 cm

\begin{figure}
\caption {Average sum $\displaystyle{ < Z_{1} + Z_{2} > }$ plotted as a
function of $\displaystyle{  Z_{1}}$ for $\displaystyle {^{35}{\rm Cl} +
{^{24}{\rm  Mg}}}$. The solid line shows EHFM+CASCADE results.}
\label {FIG.3 :}
\end{figure}

%
%

\end{document}